\documentclass[a4paper,twoside]{article}
\newcommand{\affil}[1]{$^{\rm #1}$}
\usepackage[authoryear]{natbib}
\usepackage{lscape}
\bibpunct{(}{)}{;}{a}{}{,}
\usepackage{graphicx}
\date{} %Please leave the date blank
\newcommand{\arcsec}{\mbox{$^{\prime\prime}$}}
\newcommand{\arcmin}{\mbox{$^\prime$}}
\title{\large\bf\flushleft Ks-band (2.14$\mu$m) imaging of southern
  massive star formation regions traced by methanol masers}

\author{\parbox{\textwidth}{\flushleft
\vspace{-0.5cm}
{\it S. N. Longmore\affil{A,B,C,D} and M. G. Burton\affil{A}}\\
\vspace{0.4cm}
{\small \affil{A}\,School of Physics, University of New South Wales, Sydney, NSW 2052, Australia}\\
{\small \affil{B}\,Australia Telescope National Facility, CSIRO, Epping 1710, Australia}\\
{\small \affil{C}\,Present address: Harvard-Smithsonian Center for Astrophysics, Cambridge, MA 02138, USA}\\
{\small \affil{D}\,Email: slongmore@cfa.harvard.edu}}}
\begin{document}

\twocolumn[
\begin{changemargin}{.8cm}{.5cm}
\begin{minipage}{.9\textwidth}
\vspace{-1cm}
\maketitle
%
%%%%%%%%%%%%%     ABSTRACT    %%%%%%%%%%%%%
%Abstract of no more than 200 words here.
\small{\bf Abstract:}

We present deep, wide-field, Ks-band (2.14 $\mu$m) images towards 87
southern massive star formation regions traced by methanol maser
emission. Using point-spread function fitting, we generate 2.14$\mu$m
point source catalogues (PSCs) towards each of the regions. For the
regions between 10$^\circ<l<$~350$^\circ$ and $|b|<$1, we match the
2.14 $\mu$m sources with the GLIMPSE point source catalogue to
generate a combined 2.14 to 8.0$\mu$m point source catalogue. We
provide this data for the astronomical community to utilise in studies
of the stellar content of embedded clusters.

%%%%%%%%%%%%%     KEYWORDS    %%%%%%%%%%%%%
\medskip{\bf Keywords:} stars: formation, stars: early type,
infrared: stars, Galaxy: stellar content, masers
% Please write all keywords in lower case. PASA uses the
% standard list of subject headings adopted by The Astrophysical Journal
% and available from http://www.journals.uchicago.edu/ApJ/keywords_text.html.
% Keywords are separated by em-dashes, i.e. ---

%%%%%%%%DO NOT EDIT%%%%%%%%%%%%
\medskip
\medskip
\end{minipage}
\end{changemargin}
]
\small

%%%%%%%%EDIT FROM HERE%%%%%%%%%%%%
\section{Introduction}

Stellar populations towards young star formation regions are deeply
embedded within large column densities of molecular gas and dust.
Observations at IR wavelengths are particularly powerful for peering
into these natal cocoons as they suffer much less from extinction than
optical wavelengths and the young stars may also be intrinsically
cool. Several excellent large-area IR surveys exist [e.g. 2MASS
  \citep{skrutskie2006}, GLIMPSE \citep{benjamin2003}], the results
from which have significantly advanced our understanding of early
phases of cluster formation. However, these surveys are limited to
$\sim$2$\arcsec$ resolution which may be insufficient to resolve the
closest cluster members \citep[e.g.][]{longmore2006} and the 2MASS
Ks-band (2.14\,$\mu$m) observations are not deep enough to detect the
most heavily embedded/reddest objects seen in the GLIMPSE images.

In this work we present 2.14$\mu$m images and photometry for 87 fields
associated with massive star formation regions. These were originally
targetted due to their IRAS colours in a search for methanol maser
emission \citep{walsh1997,walsh1998} and most contain hot molecular
cores \citep{L07A,purcell2006,purcell2009}. By matching source
positions with those in existing near-IR (2MASS) and mid-IR (GLIMPSE)
catalogues, we have created composite 2.14 to 8\,$\mu$m catalogues
which are typically 3-4 magnitudes deeper at 2.14$\mu$m than
2MASS. Detailed analysis of these regions requires consideration of
completeness and confusion etc., which is specific to individual
fields -- a result of differing levels of differential extinction and
source confusion as well as differing sensitivity across the
fields. However, we calculate a number of metrics which describe the
quality and quantity of the data sets. We provide these catalogues and
calibrated images as a resource for the astronomical community.

%------------------------------------------------------------
\section{Observations and Data Reduction}
\label{sec:iris2_obs}
The observations were made using IRIS2\footnote{IRIS2 employs a
  1024x1024 Rockwell HAWAII-1 HgCdTe infrared detector with a plate
  scale of 0.4486 $\pm$ 0.0002$\arcsec$ per pixel.}$^,$\footnote{see
  http://www.aao.gov.au} (Infrared Imager and Spectrometer 2) on the
3.9m Anglo Australian Telescope (AAT) at Siding Spring Observatory
using the Ks-band (1.982--2.306\,$\mu$m) filter. 87 regions were
imaged, as listed in Table~\ref{tab:daophot_pars}. The images are
shown in Figures 7 to 92. The observations were taken over several
different periods between 2003 and 2006. For each source, a 3 $\times$
3 image grid was created with $\sim$ 1$\arcmin$ offsets from the
pointing centre. The integration time at each of the nine grid
positions was around one minute, with 9$\times$6s exposures. The
initial reduction was carried out using a data reduction pipeline
provided by the Anglo-Australian Observatory. The detector bias was
removed at readout time using the Double Read Mode method. The data
were reduced using the in-house ORAC-DR pipeline with the
`JITTER\_SELF\_FLAT\_KEEPBAD' ORAC-DR\footnote{see
  http://www.oracdr.org} recipe to correct for dark current, create
flat-field images from star free pixels in the 9 jittered source
fields and apply a bad pixel mask to each image. The pipeline then
corrects for distortion at the outer edges of the image caused by the
slightly curved focal plane. The 9 individual fields of
7.7$\arcmin$$\times$7.7$\arcmin$ were finally aligned and mosaiced
together to give a 9.7$\arcmin$$\times$9.7$\arcmin$ field of view,
although only the inner 5.7$\arcmin \times$5.7$\arcmin$ are covered by
all 9 fields. A typical sensitivity in the inner region of an
uncrowded field would be expected to be $\sim$19.6\,mags (5$\sigma$)
in good conditions.

\subsection{Co-ordinate Frame Calibration}
\label{sub:pointing}
The blind telescope pointing error is estimated to be $\sim$3$\arcsec$
but absolute coordinates were calculated by comparing non-saturated
stars to those in the 2MASS catalogue. Using the \emph{koords} program
in the
KARMA\footnote{http://www.atnf.csiro.au/computing/software/karma/}
visualisation package, the image coordinates were matched to better
than 0.1 pixel accuracy. We take the error in the absolute coordinates
to be the error in the 2MASS catalogue of 0.3$\arcsec$.

\subsection{Source Extraction}
\label{sub:source_extract}

\begin{figure*}
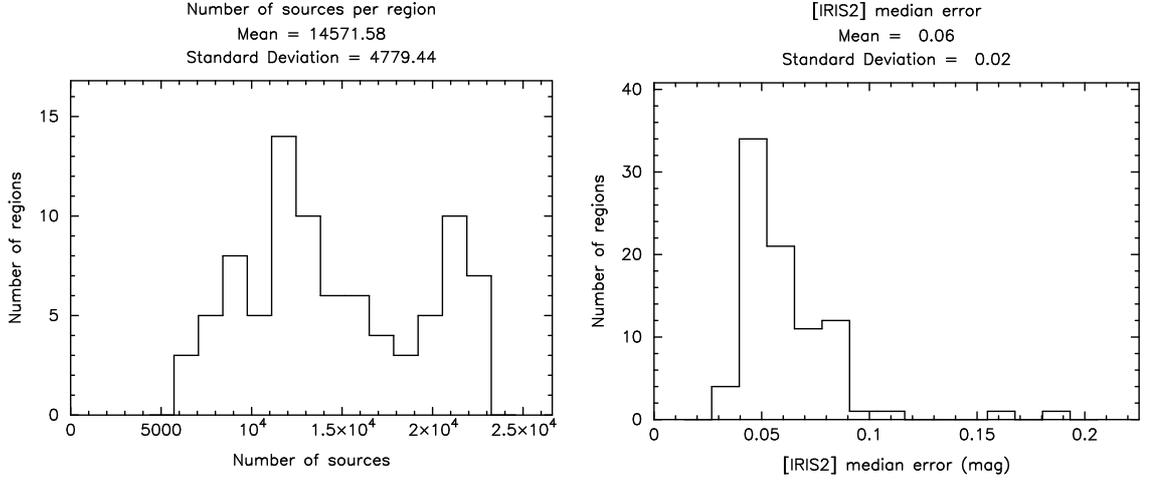

\begin{center}
\begin{tabular}{cc}
 \includegraphics[height=0.45\textwidth, angle=-90, trim=0 0 -5 0]{figs/ps_files/plot_num_source_hist.ps} &
 \includegraphics[height=0.45\textwidth, angle=-90, trim=0 0 -5 0]{figs/ps_files/plot_med_err_mag_hist.ps}\\
\end{tabular}
\caption{[Left] Histogram of the number of sources detected per region
  and [Right] the median source photometric fit uncertainty per region
  determined using DAOPHOT.}
\label{fig:num_src_per_reg}
\end{center}
\end{figure*}

Sources were extracted from each of the IRIS2 images using the
\emph{daophot} tasks in the IRAF\footnote{http://iraf.noao.edu/}
package to produce a 2.14\,$\mu$m source catalogue (see
Section~\ref{sub:creating_PSCs} for more details). Examination of the
residuals after removing the stellar flux from the image shows the
stars have generally been well extracted despite different background
levels across the image (caused by varying extended emission and
extinction) and the crowded field. Figure~\ref{fig:num_src_per_reg}
shows a histogram of the number of sources detected per region (which
varied between $\sim$6,000 to 25,000, corresponding to stellar
densities $\sim$60 $-$ 250 stars/arcmin$^2$) and the median source
photometric fit uncertainty per region ($\sim$0.06\,mag).

\subsection{Photometric Calibration}
\label{sub:photometry}
\begin{figure*}
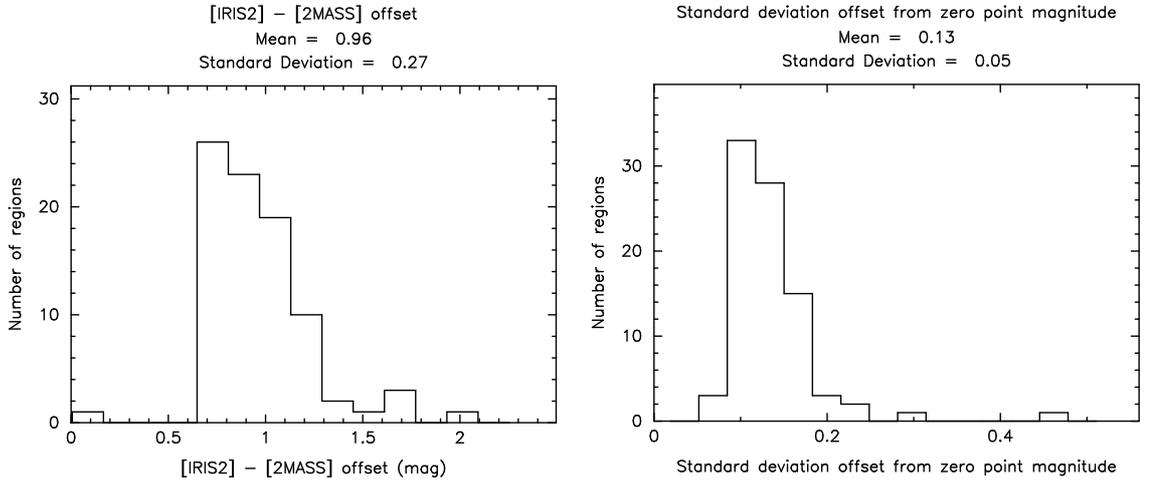

\begin{center}
\begin{tabular}{cc}
 \includegraphics[height=0.45\textwidth, angle=-90, trim=0 0 -5 0]{figs/ps_files/plot_zp_hist.ps}&
 \includegraphics[height=0.45\textwidth, angle=-90, trim=0 0 -5 0]{figs/ps_files/plot_d_2mass_i2_hist.ps}\\
\end{tabular}
\caption{Histograms illustrating photometric calibration outlined in
  Section~\ref{sub:photometry}. [Left] Histogram of the
  [IRIS2]$-$[2MASS] instrumental magnitude offset for each region,
  used to calibrate the IRIS2 catalogues. [Right] Histogram of the
  estimated photometric calibration accuracy towards each region. This
  was derived from the standard deviation between IRIS2 and 2MASS
  magnitudes of sources over a reliable magnitude range
  ($\sim$12$^{\rm th}$ to 14$^{\rm th}$ magnitude) after the
  calibration factor had been applied. }
\label{fig:i2_photometry}
\end{center}
\end{figure*}

The photometry was calibrated using the 2MASS catalogue towards each
region (see Section~\ref{sub:2mass}). The extracted IRIS2 sources
(Section~\ref{sub:source_extract}) were first matched with the 2MASS
catalogue as outlined in Section~\ref{subsec:match_catags}. The mean
offset between the two catalogues was then calculated by plotting the
difference between the 2MASS magnitude and IRIS2 instrumental
magnitude of the matched stars as a function of the measured IRIS2
instrumental magnitude over a suitable magnitude range (typically
$\sim$12 to 14, to avoid both IRIS2 saturation and the poorer 2MASS
sensitivity). Figure~\ref{fig:i2_photometry} shows a histogram of the
offset between the IRIS2 instrumental magnitude and 2MASS magnitude
for each region. The range in values is due to the different airmass
of the IRIS2 observations. The uncertainty in IRIS2 absolute
photometry was estimated from the standard deviation between IRIS2 and
2MASS magnitudes over a reliable magnitude range ($\sim$12$^{\rm th}$
to 14$^{\rm th}$ magnitude) after this offset had been
applied. Figure~\ref{fig:i2_photometry} shows a histogram of the
estimated photometric accuracy towards each region, which is
$\sim$0.13 magnitudes averaged over all the fields. This is
approximately twice the statistical error from the photometric fitting
(see Figure~\ref{fig:num_src_per_reg}). Some of this scatter is likely
to be due to the difference between the MKO photometric system used by
IRIS2 and the 2MASS photometric
system\footnote{http://www.astro.caltech.edu/$\sim$jmc/\\\hspace{2cm}2mass/v3/transformations/}
\citep{carpenter2001}. Of particular relevance to this work, the Ks
filters of the two photometric systems are responsive over different
wavelength ranges -- from 1.982 to 2.306\,$\mu$m for IRIS2 and 1.915
to 2.384\,$\mu$m for 2MASS. Although converting to the 2MASS
photometric system may marginally reduce this scatter, the colour
correction terms are not characterised for sources with (J-H) $>$ 1.5,
(H-K) $>$ 1 and (J-K) $>$ 2. These are the reddest, most embedded
sources and thus likely to be of most interest to studies of star
formation.

\begin{figure*}
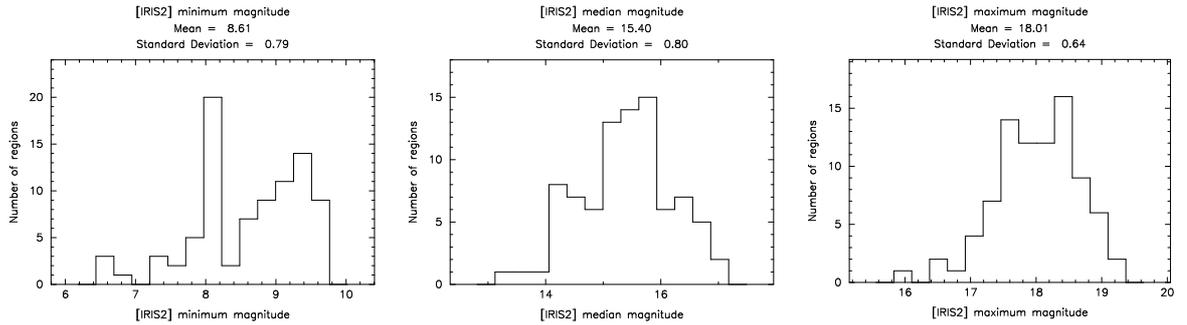

\begin{center}
\begin{tabular}{ccc}
 \includegraphics[height=0.3\textwidth, angle=-90, trim=0 0 -5 0]{figs/ps_files/plot_minmag_hist.ps}&
 \includegraphics[height=0.3\textwidth, angle=-90, trim=0 0 -5 0]{figs/ps_files/plot_medmag_hist.ps}&
\includegraphics[height=0.3\textwidth, angle=-90, trim=0 0 -5 0]{figs/ps_files/plot_maxmag_hist.ps} \\
\end{tabular}
\caption{Histograms of brightest, median and faintest magnitude of
  sources detected per region.}
\label{fig:min_med_max_mag}
\end{center}
\end{figure*}

Figure~\ref{fig:min_med_max_mag} shows histograms of brightest, median
and faintest magnitude of sources detected per region. Depending on
the field, stars brighter than $\sim$11$^{\rm th}$ magnitude in the
IRIS2 fields suffer from saturation. Column 11 in
Table~\ref{tab:daophot_pars} gives the minimum reliable magnitude for
each region, corresponding to the maximum flux per field over which
the array responded linearly. Photometry for stars brighter than this
should be taken from the 2MASS catalogue.

Table~\ref{tab:daophot_pars} also provides information on the
extracted IRIS2 2.14$\mu$m source populations, image quality and
photometric calibration towards each of the regions. The photometric
data themselves are provided in an online catalogue in
Table~\ref{tab:i2ks_psc_examp}. We provide here the first five lines
and a description of the PSC contents. These are further combined with
the photometry for matched sources in the GLIMPSE catalogue (see
Section~\ref{subsec:match_catags}). The first five lines of the
combined catalogue are provided here, with the rest in the online
material.

%------------------------------------------------------------
\subsection{Matching sources at different wavelengths}
\label{subsec:match_catags}
Analysis of source colours relies on the ability to accurately match
sources in the catalogues at different wavelengths. With too many
sources ($>$10$^4$) to check manually and crowded stellar fields,
there is a potential problem of either matching a source multiple
times or mismatching the sources completely. The resulting mismatched
sources may replicate the large colour differences of embedded sources
and hence contaminate the sample. To test the matching accuracy, we
generated and matched synthetic catalogues with known offsets from a
list of absolute source positions. In order to ensure a realistic
spatial distribution of sources, including over densities due to
clustering and under densities due to extinction, the absolute
coordinates were taken from the 2MASS PSC of the region. The
coordinates recorded in the synthetic catalogues were randomly offset
from the absolute values by a 2D Gaussian distribution. Several
synthetic catalogues were produced using the known absolute pointing
error of the relevant datasets to generate the positional offsets. The
sources in the synthetic catalogues were given a unique ID and then
matched in the same way as the observed datasets using the
\emph{tmatch} routine in the \emph{tables.ttools} package in IRAF.
The routine matches every source in the first catalogue with the
nearest counterpart in the second catalogue within a user defined
matching radius. Comparing the ID of the matched sources in the
synthetic catalogues it is possible to unambiguously identify which
sources have been matched correctly.

The IRIS2 images were all individually registered with 2MASS images so
have a \emph{relative} astrometric uncertainty of
$\Delta_{IRIS2}=0.04\arcsec$ (0.1 pixel) [see
  Section~\ref{sub:pointing}].  The astrometric uncertainty in the
2MASS and GLIMPSE surveys are both quoted as 0.3$\arcsec$. However,
the GLIMPSE survey coordinates were also registered with 2MASS data so
the relative uncertainty between the IRIS2 and GLIMPSE catalogues
should be considerably less than the assumed error of
$\Delta_{IRIS2-GLIMPSE}=0.3\arcsec$.  Matching two synthetic
catalogues generated with an uncertainty of $\Delta_{IRIS2}$, an equal
number of sources and a search radius of 2 pixels, produces a 100\%
recovery with no confusion due to doubles or mismatches. Matching
catalogues generated with an uncertainty of $\Delta_{IRIS2}$ but
different numbers of sources also recovers 100\% of the sources
correctly but includes $\sim$0.05\% double sources. With different
numbers of sources in each catalogue, the matching order then becomes
important. For example, matching the catalogue with larger to smaller
numbers of sources may produce `multiple matches', where several
sources in the larger catalogue are within the matching radius of a
source in the smaller catalogue. However, the number of correctly
matched sources is the same, irrespective of the matching
order. Varying the matching radius has little effect as long as it is
significantly larger than the combined astrometric uncertainty.

Based on the above analysis, when matching the IRIS2 PSC with the
GLIMPSE PSC and a matching radius of 1.2$\arcsec$, we expect to
recover at least 99.8\% of the sources correctly with a maximum of
0.1\% doubles and 0.12\% mismatches. We therefore conclude that
mismatches should not seriously affect the combined IRIS2/GLIMPSE
PSCs.

%------------------------------------------------------------
\section{Other Catalogue Data}
\subsection{2MASS - 2 micron All Sky Survey}
\label{sub:2mass}
Using two 1.3m telescopes in Arizona and Chile, the 2MASS survey
scanned the entire sky at the J, H and Ks bands (1.25, 1.65 and
2.17$\mu$m) \citep{skrutskie2006}. The survey has a 10$\sigma$
limiting magnitude of 15.8, 15.1, 14.3 $\pm$ 0.03 mag at J, H and Ks
respectively with a pixel size of 2.0$\arcsec$ and 0.1$\arcsec$
pointing accuracy. We make use of this data to provide a reference
frame for the images as well as to provide the flux calibration.

\subsection{GLIMPSE - Galactic Legacy Infrared Mid-Plane Survey Extraordinaire}
\label{sub:glimpse}
Using the infrared array camera (IRAC) on-board the Spitzer Space
Telescope (SST\footnote{http://ssc.spitzer.caltech.edu/}), the GLIMPSE
survey has observed the galaxy at 3.6, 4.5, 5.8 and 8.0$\mu$m with a
1.2" pixel size between $10^\circ<|l|<65^\circ$ and $|b|<1^\circ$
\citep{benjamin2003}.  The photometric accuracy is 0.2 and 0.3 mag for
bands 1/2 and 3/4 respectively. The astrometric accuracy of the point
source catalogue is $\sim$0.3$\arcsec$. We combine this with the AAT
data to construct a source catalogue from 2.14 to 8.0$\mu$m.

%------------------------------------------------------------
\section{The data}
\label{sec:data}

Figures~7~to~92 show the IRIS2 2.14\,$\mu$m images towards each of the
regions overlayed with the methanol maser emission from
\citet{walsh1998}. Note that in most cases, regions of extinction are
readily apparent. Rarely is there a 2$\mu$m point source associated
with a methanol maser. In many cases, extensive regions of 2$\mu$m
nebulosity are seen. Several targetted regions overlap within a few
arcminutes (e.g. G14.99-0.70 and G15.03-0.68 in the vicinity of M17)
but each field was imaged independently to ensure uniform coverage
(within the observational systematics) in the 5.7$^\prime \times
5.7^\prime$ inner region of the mosaic with optimum sensitivity (see
Section~\ref{sec:iris2_obs}). 

Table~\ref{tab:i2ks_psc_examp} shows an example extract from the first
five lines of the IRIS2 Ks-band (2.14\,$\mu$m) point source catalogue
for a sample region. Sections~\ref{sub:seeing} \&
\ref{subsec:completeness} examine the issues of seeing and
completeness in these Ks-band (2.14\,$\mu$m) PSCs including photometry
for these sources. 

The fits images, IRIS2 PSCs and IRIS2/GLIMPSE PSCs for each region are
available through the Centre de Donn\'{e}es astronomiques de
Strasbourg (CDS)\footnote{http://cds.u-strasbg.fr/}. Photometric
calibration for the fits images is provided by the fits header
keyword, ``MAGZERO'', which gives the magnitude in each image
corresponding to a flux count of 1\footnote{i.e. m = m$_{ZP}$ - 2.5
  log(f), where f is the number of counts determined for a source of
  interest in the fits images}. This is the same value as that given
in column 6 of Table~\ref{tab:daophot_pars}.

\subsection{Seeing}
\label{sub:seeing}

\begin{figure*}
\begin{center}
\begin{tabular}{ccc}
 \includegraphics[height=0.3\textwidth, angle=-90, trim=0 0 -5 0]{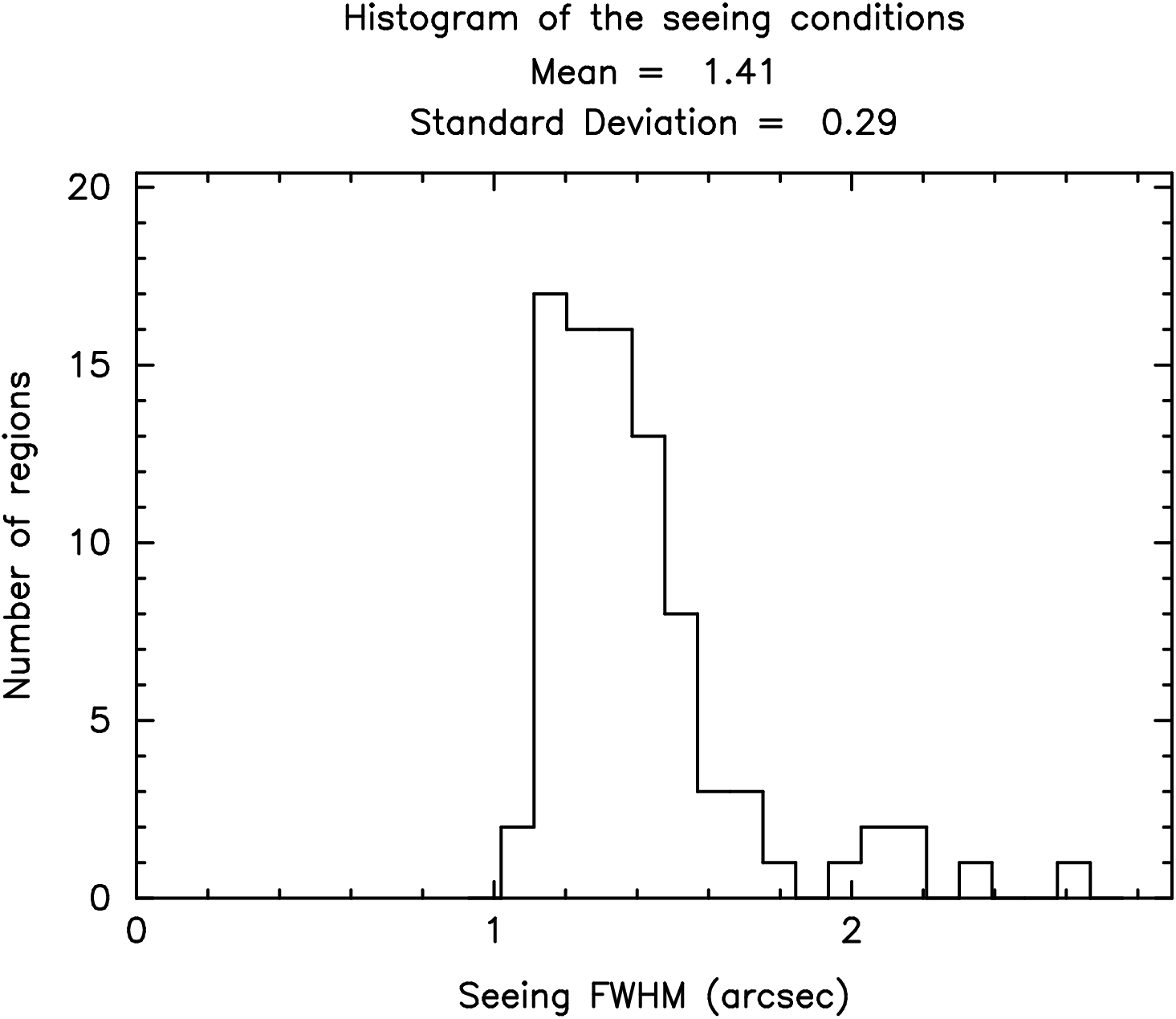}&
 \includegraphics[height=0.33\textwidth, angle=-90, trim=0 0 -5 0]{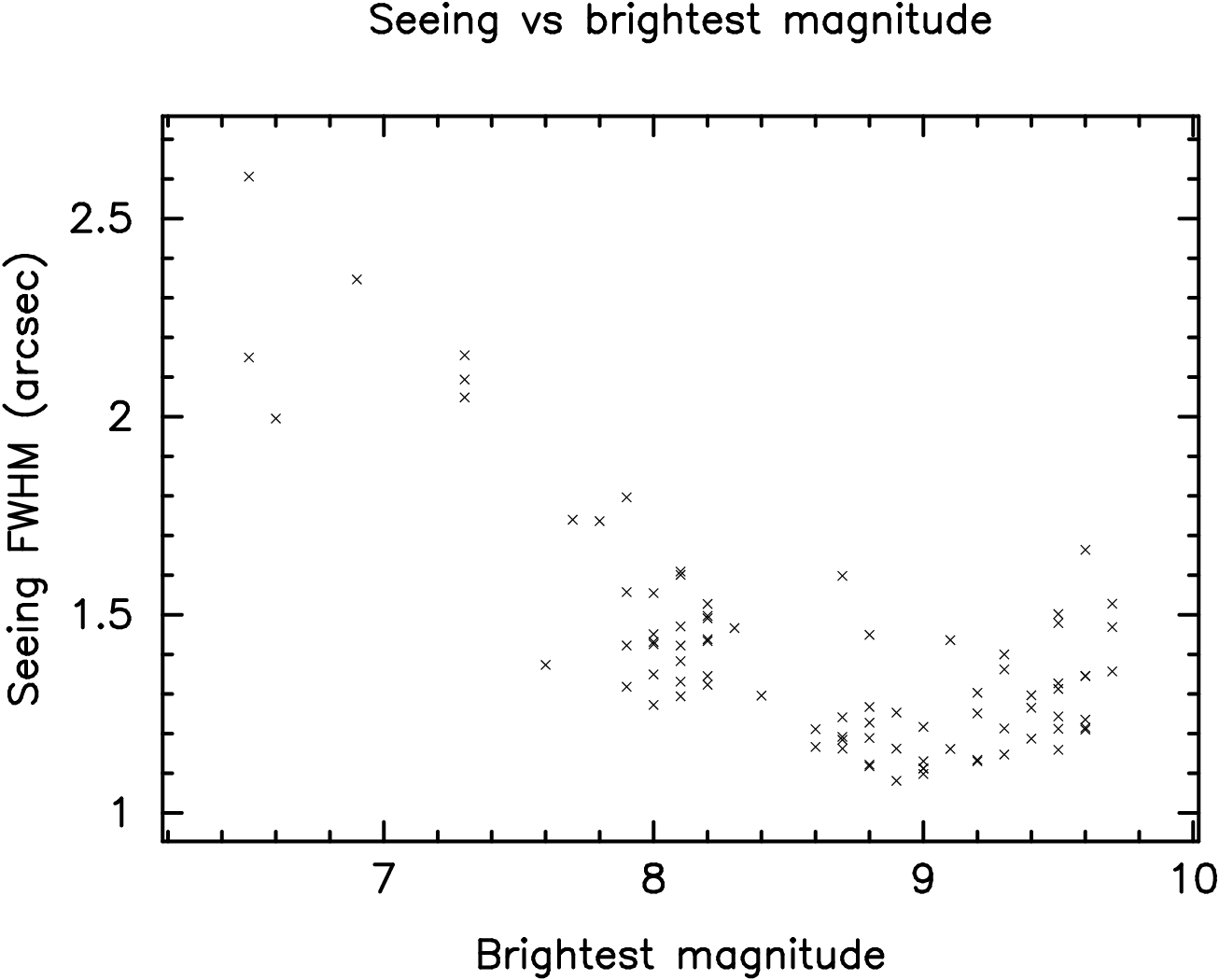}&
\includegraphics[height=0.33\textwidth, angle=-90, trim=0 0 -5 0]{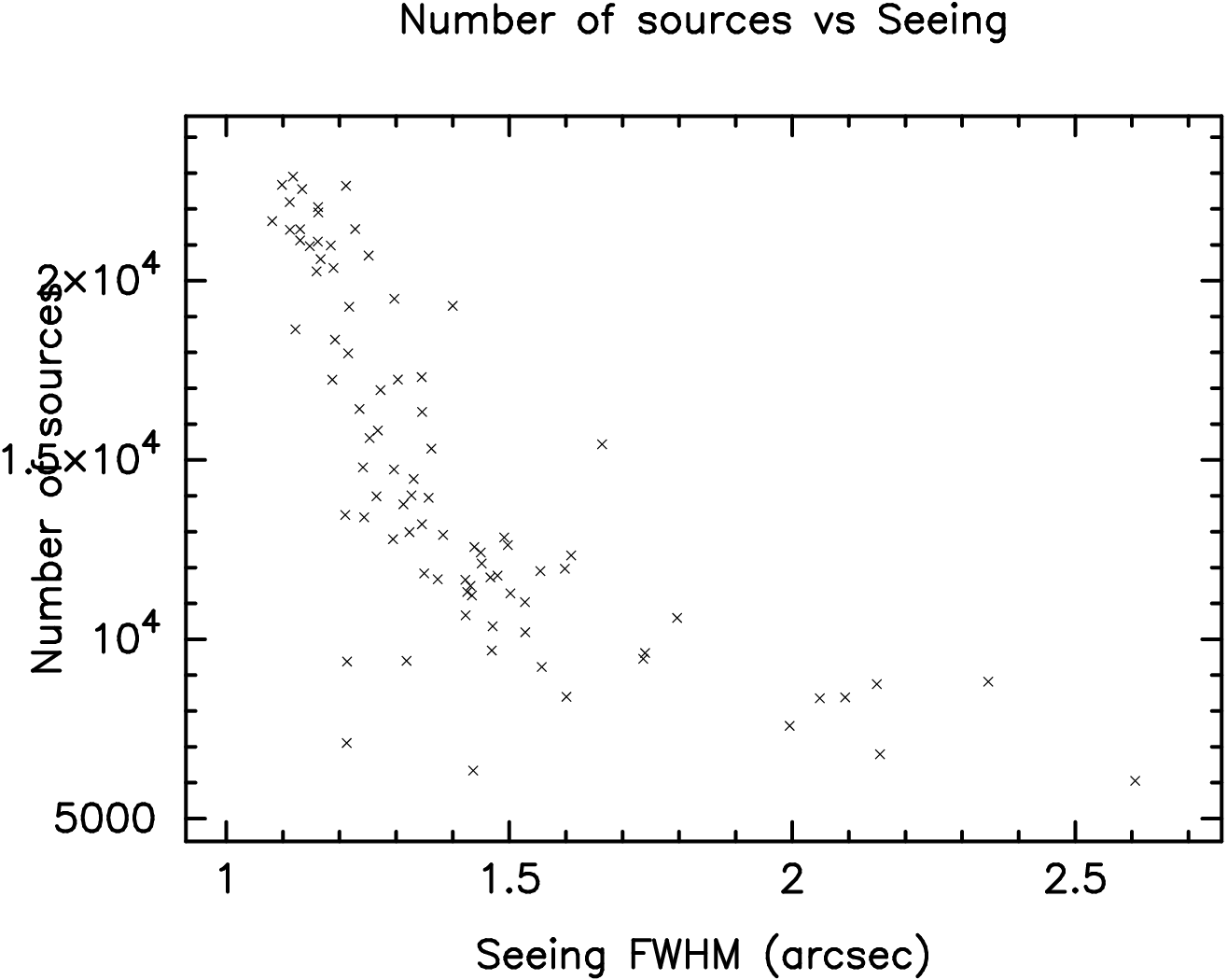} \\

\end{tabular}
\caption{[Left] Histogram of seeing conditions during observations of
  the regions, [Centre] brightest detected magnitude per region vs
  seeing and [Right] number of sources detected per region vs seeing.}
\label{fig:seeing}
\end{center}
\end{figure*}

Figure~\ref{fig:seeing} shows a histogram of the seeing conditions
during observations of each of the regions. The majority of regions
were observed in seeing of 1.2 to 1.4 arcseconds, a noticeable
improvement on the $\sim$2$\arcsec$ resolution of both 2MASS and
GLIMPSE. The other two plots in Figure~\ref{fig:seeing} show the
effects of seeing on the brightest magnitude and number of sources
detected. When the seeing is poor, the flux from bright stars is
spread over more pixels and brighter stars can be recovered. Given the
crowded nature of the observed fields towards the Galactic plane, it
is unsurprising that the number of sources detected increases
dramatically as the seeing decreases. The inverse linear relation
between source counts and seeing suggests the difference in source
counts resulting mostly due to the seeing present at the time of each
observation.

\subsection{Completeness}
\label{subsec:completeness}

\begin{figure}
\begin{center}
 \includegraphics[height=0.4\textwidth, angle=-90, trim=0 0 -5 0]{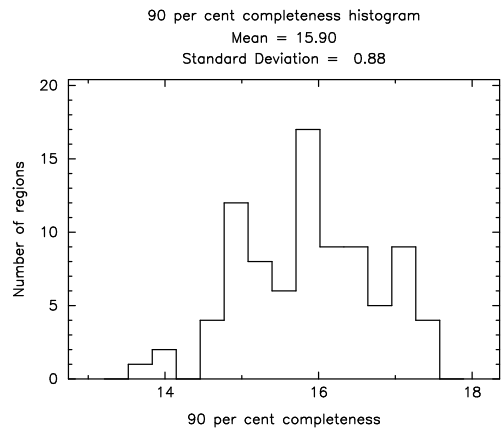}

\caption{Histogram of estimated 90\% completeness using a simple,
  empirical method of calculating the turnover in the histogram of the
  number of stars as a function of magnitude, following L07B.}
\label{fig:completeness}
\end{center}
\end{figure}

\citet[][hereafter L07B]{L07B} used artificial star recovery to
investigate the spatial variation in point source sensitivity as a
function of wavelength for the field associated with
G305.2+0.2. Having calculated the PSF for each image, a grid of
artificial stars of the same magnitude separated by 30$\arcsec$ was
placed across the image. The same automated finding technique outlined
in Section~\ref{sec:iris2_obs} was then used to calculate how many of
the artificial stars were recovered. By shifting the
30$\arcsec\times$30$\arcsec$ grid of artificial stars in small steps
through the image, it was possible to measure the completeness at
5$\arcsec$ intervals without the PSF of individual artificial stars
overlapping. The process was repeated by increasing the artificial
star magnitudes in steps of 0.5 mag until no more stars were
recovered. By recording the largest recovered magnitude at each
position and wavelength it was possible to build a three-dimensional
picture of the point source sensitivity across the region. The
\emph{relative} completeness as a function of position was then
calculated at each wavelength by subtracting the median completeness
magnitude at that wavelength from every position. This method is
similar to that used by \citet{gutermuth2005} in an analysis of
completeness limits of Spitzer IRAC data. Through analysis of the
results from the artificial star recovery method, L07B found a
simpler, empirical method of estimating the 90\% completeness limit by
calculating the turnover in the histogram of the number of stars as a
function of magnitude which yields approximately the same limit as the
full analysis. This metric is provided in Table 1, but we stress that
a comprehensive analysis of each field would require this completeness
limit to be determined individually for
each. Figure~\ref{fig:completeness} presents the histogram of this
distribution of turnover magnitude.

%------------------------------------------------------------

\section{Applications}
\begin{figure}
\begin{center}
%\begin{tabular}{ccc}
 \includegraphics[width=7.5cm, height=7.5cm, angle=-90, trim=0 0 -5 0]{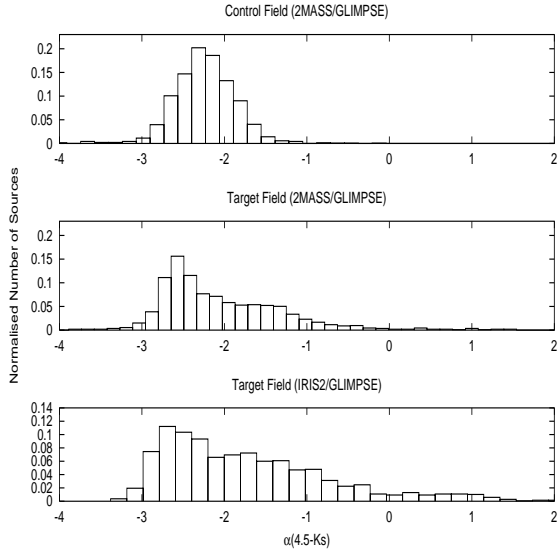}
%\end{tabular}
\caption{Histogram of the distribution of spectral indices
  ($\alpha=~d[log(\lambda F_\lambda)]/d[log(\lambda)]$) for all
  sources matched at Ks (2.14\,$\mu$m) and 4.5$\mu$m towards the region
  G305.2+0.2 \citep[from][]{L07B}. The top and central figures show
  the distribution in the control field and target region
  respectively, for sources matched in the 2MASS/GLIMPSE PSC. The
  bottom figure gives the distribution of sources matched in the
  IRIS2/GLIMPSE PSC of the target region. The considerably more
  sensitive 2.14\,$\mu$m IRIS2 images means much more deeply embedded
  (redder) sources are detected, which is seen in the bottom figure as
  a significant increase in the number of sources with $\alpha \geq
  0$.}
\label{fig:a_k_4.5_hist}
\end{center}
\end{figure}

As an example of what the data has the potential to provide once a
full completeness analysis is conducted, we show the results towards
G305.2+0.2, as reported by L07B. Figure~\ref{fig:a_k_4.5_hist} shows
the slope of the spectral energy distribution between 2.14 and
4.5\,$\mu$m (i.e. $\alpha_{2.14-4.5}$) of all sources detected towards
i) a control field using a combined 2MASS/GLIMPSE PSC, ii) the target
field using a combined 2MASS/GLIMPSE PSC and iii) the target field
using a combined IRIS2/GLIMPSE PSC. The spectral index is related to
the categories of young stellar objects (Class I, II, III), with
younger sources tending towards larger spectral indexes. In this way
the PSC provides a measure of the distribution of evolutionary states
of the embedded cluster seen towards G305.2+0.2. The considerably more
sensitive 2.14\,$\mu$m IRIS2 images means much more deeply embedded
(redder) sources are detected with the IRIS2/GLIMPSE PSC compared to
the 2MASS/GLIMPSE PSC. This is obvious in the bottom panel of
Figure~\ref{fig:a_k_4.5_hist} as a significant increase in the number
of sources with $\alpha \geq 0$ compared to the middle panel.

\section{Acknowledgements}
We thank Stuart Ryder for help with the data reduction. SNL was
partially funded by an internal grant from UNSW.

\bibliography{iris2_survey_astroph}

\section{Appendix}
Table~\ref{tab:daophot_pars} shows the properties of the IRIS2
2.14\,$\mu$m PSCs for all the observed
regions. Tables~\ref{tab:i2ks_psc_examp}~\&~\ref{tab:i2ks_glimpse_psc_examp}
show the first five lines of a sample IRIS2 2.14\,$\mu$m PSC and a
sample combined IRIS2/GLIMPSE PSC, respectively, with the full tables
available on line. Finally, Section~\ref{sub:creating_PSCs} describes
in detail how the DAOPHOT package (digi.daophot) in IRAF was used to
automatically find and perform photometry on stars within the IRIS2
fields and hence create the point source catalogues.

 \begin{table*}
  
    \caption{Summary of the data characteristics for each of the
      regions observed with IRIS2. Columns one and two give the
      Galactic and IRAS names of the far-IR sources whose fields have
      been imaged. Column three and four give the R.A.~and Dec.~of the
      associated methanol masers from \citet{walsh1998}. Columns five
      to thirteen show the parameters returned from the automated
      source extraction algorithm using \emph{DAOPHOT} (see
      Section~\ref{sub:creating_PSCs} for more details). $\sigma_{ZP}$
      gives the standard deviation of the difference between the
      measured IRIS2 and 2MASS magnitudes (in the range 10$-$12 mags).
      ZP gives the magnitude corresponding to 1 flux count in each
      image. This may be used to calibrate the fits image for each
      source, as described in Section~\ref{sec:data}. Min, Med and Max
      give the minimum, median and maximum magnitudes of sources in
      each region. $\sigma_{med}$ gives the median point-source
      fitting error in each region. Min$_{rel}$ gives the minimum
      reliable IRIS2 magnitude measurement, sources brighter than
      which are likely to be affected by saturation. Columns twelve
      and thirteen show the 90\% completeness magnitude estimated from
      the magnitude histogram and the total number of sources detected
      in each IRIS2 field, respectively. The final column lists the
      FWHM of the seeing in arcsec. $^a$The dashes in column 2 show
      sources with no associated IRAS point source. $^b$G305.20+0.21
      has been included in this table for completeness but the
      catalogue and images can be found in \citet{L07B}.}
    \label{tab:daophot_pars}
    \vspace{2mm}
    \begin{tiny}

    \begin{tabular}{lcccccccccccccccc}\hline
 Region          & IRAS$^a$    &RA           & Dec         &   \multicolumn{8}{c}{Magnitudes}  & Num & FWHM   \\
                 &             &(J2000)      & (J2000)     & $\sigma_{ZP}$ & ZP  & Min.  & Med.  & Max.  & $\sigma_{med.}$ & Min$_{rel}$ & 90\% & sources & (arcsec) \\ \hline

G290.40-2.91     & 10555-6242  & 10:57:34.0 & -62:59:3.6  & 0.11 & 22.95 & 8.1 & 17.0 & 18.6 & 0.16 & 11.0 & 17.5 & 8395 &  1.60\\
G293.82-0.74     & 11297-6156  & 11:32:05.4 & -62:12:25.0 & 0.16 & 23.66 & 9.3 & 16.4 & 17.8 & 0.19 & 10.0 & 16.9 & 9380 &  1.21\\ 
G305.20+0.21$^b$ & 13079-6218  & 13:11:10.5 & -62:34:38.9 & 0.20 & 24.01 & 5.3 & 16.1 & 18.5 & 0.06 & 11.0 & 17.5 & 11226 & 1.20 \\
G309.92+0.48     & 13471-6120  & 13:50:41.8 & -61:35:10.1 & 0.09 & 24.01 & 9.7 & 16.3 & 18.5 & 0.09 & 11.0 & 16.9 & 9689 &  1.47\\
G312.11+0.26     & 14050-6056  & 14:08:49.3 & -61:13:26.0 & 0.09 & 24.12 & 9.6 & 16.6 & 19.0 & 0.09 & 11.0 & 17.3 & 15439 &  1.66\\
G313.77-0.86     & 14212-6131  & 14:25:01.6 & -61:44:58.1 & 0.08 & 24.20 & 9.5 & 15.9 & 18.5 & 0.06 & 11.0 & 16.9 & 11775 &  1.48\\
G316.81-0.06     & 14416-5937  & 14:45:26.9 & -59:49:16.3 & 0.16 & 24.24 & 8.7 & 16.3 & 18.7 & 0.08 & 11.0 & 16.9 & 14796 &  1.24\\
G318.95-0.20     & 14567-5846  & 15:00:55.3 & -58:58:53.6 & 0.22 & 24.11 & 9.5 & 16.5 & 18.8 & 0.09 & 12.0 & 17.1 & 20259 &  1.16\\
G320.23-0.28     & 15061-5814a & 15:09:52.0 & -58:25:38.1 & 0.07 & 24.11 & 9.3 & 16.3 & 18.8 & 0.08 & 11.0 & 16.5 & 19299 &  1.40\\
G321.03-0.48     & 15122-5801  & 15:15:51.6 & -58:11:17.4 & 0.14 & 24.09 & 9.7 & 15.3 & 17.8 & 0.04 & 11.0 & 15.6 & 10195 &  1.53\\
G323.74-0.26     & 15278-5620  & 15:31:45.8 & -56:30:49.9 & 0.13 & 24.19 & 8.8 & 16.6 & 18.9 & 0.09 & 11.0 & 17.0 & 20356 &  1.19\\
G327.12+0.51     & 15437-5343  & 15:47:32.7 & -53:52:38.5 & 0.10 & 24.15 & 9.5 & 15.7 & 18.1 & 0.05 & 11.0 & 16.0 & 11281 &  1.50\\
G327.40+0.44     & 15454-5335  & 15:49:19.5 & -53:45:13.9 & 0.06 & 24.09 & 9.7 & 15.8 & 18.1 & 0.06 & 11.0 & 16.6 & 13947 &  1.36\\
G331.28-0.19     & 16076-5134  & 16:11:26.9 & -51:41:56.6 & 0.14 & 24.19 & 9.6 & 16.2 & 18.6 & 0.07 & 11.0 & 17.0 & 17974 &  1.22\\
G332.73-0.62     & 16158-5055  & 16:20:02.7 & -51:00:32.0 & 0.09 & 24.95 & 9.6 & 15.9 & 18.5 & 0.07 & 12.0 & 16.0 & 16424 &  1.24\\
G337.71-0.05     & 16348-4654  & 16:38:29.6 & -47:00:35.7 & 0.11 & 23.93 & 7.9 & 14.6 & 17.7 & 0.04 & 10.0 & 15.0 & 10598 & 1.80\\
G348.71-1.04     & 17167-3854  & 17:20:04.0 & -38:58:30.0 & 0.31 & 24.22 & 9.6 & 15.8 & 18.4 & 0.06 & 11.0 & 16.0 & 13465 & 1.21\\
G0.21+0.00       & 17430-2844  & 17:46:07.7 & -28:45:20.0 & 0.17 & 24.08 & 9.5 & 13.2 & 15.9 & 0.04 & 11.0 & 13.6 & 14013 &  1.33\\
G0.26+0.01       & -           & 17:46:11.4 & -28:42:48.0 & 0.15 & 24.07 & 9.6 & 13.6 & 16.6 & 0.04 & 11.0 & 14.1 & 17314 & 1.35\\
G0.32-0.20       & 17439-2845  & 17:47:09.1 & -28:46:16.1 & 0.16 & 24.06 & 9.4 & 14.3 & 17.2 & 0.06 & 11.0 & 15.2 & 19494 &  1.30\\
G0.50+0.19       & 17429-2823  & 17:46:04.0 & -28:24:51.3 & 0.13 & 24.11 & 9.2 & 14.0 & 16.6 & 0.05 & 11.0 & 14.8 & 20703 &  1.25\\
G0.55-0.85       & 17470-2853  & 17:50:14.5 & -28:54:31.2 & 0.15 & 23.34 & 8.2 & 14.1 & 17.3 & 0.05 & 11.0 & 14.1 & 12628 & 1.50\\
G0.70-0.04       & 17441-2822a & 17:47:24.7 & -28:21:43.7 & 0.10 & 24.13 & 9.2 & 14.5 & 17.1 & 0.05 & 11.2 & 14.9 & 17241 & 1.30\\
G0.84+0.18       & 17436-2807  & 17:46:52.8 & -28:07:35.3 & 0.13 & 24.19 & 8.8 & 14.3 & 16.8 & 0.05 & 12.0 & 14.7 & 21439 &  1.23\\
G1.15-0.12       & 17455-2800  & 17:48:48.5 & -28:01:11.8 & 0.16 & 24.22 & 9.0 & 14.9 & 17.6 & 0.05 & 12.0 & 15.2 & 19274 &  1.22\\
G2.54+0.20       & 17480-2636  & 17:50:46.5 & -26:39:44.9 & 0.17 & 24.14 & 8.8 & 14.8 & 17.6 & 0.04 & 11.0 & 15.0 & 15821 & 1.27\\
G5.90-0.43       & 17574-2403  & 18:00:40.9 & -24:04:20.6 & 0.18 & 24.20 & 8.8 & 15.7 & 18.3 & 0.06 & 11.5 & 16.2 & 18643 &  1.12\\
G6.54-0.11       & 17577-2320a & 18:00:50.9 & -23:21:29.4 & 0.11 & 24.35 & 9.0 & 15.3 & 18.1 & 0.06 & 11.5 & 15.7 & 22674 &  1.10\\
G6.61-0.08       & 17577-2320b & 18:00:54.1 & -23:17:02.0 & 0.13 & 24.28 & 8.9 & 15.3 & 18.0 & 0.06 & 11.5 & 15.5 & 21661 &  1.08\\
G8.14+0.23       & 17599-2148  & 18:03:00.8 & -21:48:10.4 & 0.15 & 24.28 & 9.0 & 15.7 & 18.4 & 0.06 & 11.5 & 15.9 & 22195 &  1.11\\
G8.67-0.36       & 18032-2137a & 18:06:19.0 & -21:37:31.9 & 0.13 & 23.35 & 7.9 & 15.1 & 17.7 & 0.06 & 11.0 & 15.8 & 9400 &  1.32\\
G9.62+0.19       & 18032-2032  & 18:06:14.8 & -20:31:37.0 & 0.11 & 24.30 & 8.9 & 15.7 & 18.3 & 0.06 & 11.7 & 16.5 & 21904 &  1.16\\
G9.99-0.03       & 18048-2019  & 18:07:50.1 & -20:18:56.7 & 0.12 & 24.30 & 9.0 & 15.6 & 18.3 & 0.06 & 11.5 & 16.5 & 21436 &  1.13\\
G10.29-0.13      & 18060-2005d & 18:08:49.4 & -20:05:59.0 & 0.13 & 24.28 & 9.0 & 16.2 & 18.7 & 0.07 & 11.5 & 17.0 & 21420 &  1.11\\
G10.30-0.15      & 18060-2005a & 18:08:55.6 & -20:05:58.0 & 0.19 & 24.29 & 9.1 & 16.2 & 18.8 & 0.07 & 12.0 & 17.0 & 21093 &  1.16\\
G10.34-0.14      & 18060-2005b & 18:09:00.0 & -20:03:35.5 & 0.20 & 24.32 & 9.2 & 16.3 & 18.9 & 0.08 & 12.0 & 17.0 & 21122 &  1.13\\
%G10.44-0.02     & 18056-1954  & 18:08:44.9 & -19:54:38.1 & \\
G10.47+0.03      & 18056-1952b & 18:08:38.2 & -19:51:49.7 & 0.13 & 24.30 & 8.7 & 15.9 & 18.7 & 0.07 & 11.5 & 16.8 & 22054 &  1.16\\
G10.48+0.03      & 18056-1952a & 18:08:37.9 & -19:51:15.0 & 0.14 & 24.29 & 8.6 & 15.8 & 18.4 & 0.06 & 11.3 & 16.5 & 20601 &  1.17\\
G10.63-0.33      & 18075-1956b & 18:10:18.0 & -19:54:04.6 & 0.10 & 24.24 & 9.3 & 16.6 & 19.0 & 0.08 & 11.5 & 17.2 & 20965 &  1.15\\
G10.63-0.38      & 18075-1956a & 18:10:29.2 & -19:55:41.2 & 0.12 & 24.29 & 9.2 & 16.6 & 19.1 & 0.08 & 12.0 & 17.5 & 22555 &  1.13\\
G11.50-1.49      & 18134-1942  & 18:16:22.1 & -19:41:27.5 & 0.10 & 24.30 & 8.8 & 16.7 & 19.0 & 0.10 & 11.5 & 17.2 & 22905 &  1.12\\
G11.94-0.15      & 18094-1840  & 18:12:17.3 & -18:40:02.8 & 0.11 & 23.36 & 7.6 & 14.3 & 17.2 & 0.06 & 10.0 & 14.5 & 11676 &  1.37\\
G11.94-0.62      & 18110-1854  & 18:14:00.9 & -18:53:26.6 & 0.23 & 24.08 & 9.3 & 15.4 & 18.2 & 0.05 & 11.5 & 16.0 & 15317 &  1.36\\
G11.99-0.27      & 18099-1841  & 18:12:51.2 & -18:40:39.7 & 0.10 & 24.16 & 8.7 & 14.4 & 17.3 & 0.03 & 11.0 & 15.0 & 11967 &  1.60\\
G12.03-0.03      & 18090-1832  & 18:12:01.9 & -18:31:55.7 & 0.16 & 24.15 & 8.2 & 14.3 & 17.2 & 0.04 & 11.0 & 14.9 & 12573 &  1.44\\
G12.18-0.12      & 18097-1825Ab & 18:12:41.0 & -18:26:21.5 & 0.09 & 24.15 & 8.3 & 14.3 & 17.0 & 0.04 & 10.5 & 15.2 & 11727 &  1.47\\
G12.21-0.09      & 18097-1825Aa & 18:12:37.5 & -18:24:08.0 & 0.10 & 24.12 & 9.5 & 14.5 & 17.0 & 0.04 & 11.0 & 15.0 & 13765 & 1.31\\
G12.68-0.18      & 18117-1753a & 18:13:54.7 & -18:01:41.3 & 0.14 & 24.15 & 8.8 & 15.7 & 18.6 & 0.05 & 11.0 & 16.0 & 12419 & 1.45\\
G12.72-0.22      & 18112-1801  & 18:14:07.0 & -18:00:37.0 & 0.14 & 24.24 & 9.5 & 16.0 & 18.5 & 0.06 & 11.0 & 16.0 & 13400 &  1.24\\
G12.89+0.49      & 18089-1732  & 18:11:51.4 & -17:31:30.2 & 0.09 & 24.23 & 9.4 & 16.4 & 18.9 & 0.08 & 11.0 & 17.1 & 17238 &  1.19\\
G12.91-0.26      & 18117-1753b & 18:14:39.5 & -17:52:00.2 & 0.14 & 24.26 & 9.4 & 17.1 & 19.3 & 0.11 & 11.0 & 17.5 & 13986 &  1.27\\
G14.60+0.02      & 18141-1615  & 18:17:01.1 & -16:14:38.7 & 0.12 & 23.47 & 8.4 & 15.1 & 17.7 & 0.05 & 10.5 & 16.0 & 14734 &  1.30\\
G14.99-0.70      & -           & 18:20:23.1 & -16:14:43.0 & 0.18 & 24.25 & 9.1 & 14.7 & 17.6 & 0.05 & 11.0 & 15.5 & 6338 &  1.44\\
G15.03-0.68      & 18174-1612  & 18:20:24.8 & -16:11:35.4 & 0.47 & 24.30 & 9.5 & 14.7 & 17.6 & 0.05 & 12.0 & 15.5 & 7107 &  1.21\\
G16.59-0.05      & 18182-1433  & 18:21:09.1 & -14:31:48.7 & 0.12 & 24.30 & 8.7 & 15.8 & 18.5 & 0.07 & 11.0 & 16.0 & 20984 &  1.18\\
G16.86-2.16      & 18265-1517  & 18:29:24.4 & -15:16:04.1 & 0.09 & 23.92 & 8.2 & 15.8 & 18.1 & 0.07 & 10.0 & 16.2 & 11039 &  1.53\\
G19.36-0.03      & 18236-1205  & 18:26:25.2 & -12:03:52.8 & 0.11 & 24.26 & 8.7 & 15.6 & 18.4 & 0.06 & 11.5 & 15.8 & 18355 &  1.19\\ 
G19.61-0.13      & 18244-1155  & 18:27:16.4 & -11:53:38.4 & 0.09 & 24.28 & 8.6 & 15.5 & 18.3 & 0.06 & 11.5 & 15.8 & 22645 &  1.21\\
G19.70-0.27      & 18248-1158  & 18:27:55.9 & -11:52:39.1 & 0.16 & 24.23 & 8.9 & 14.6 & 17.2 & 0.05 & 11.5 & 15.0 & 15609 &  1.25\\
G21.88+0.01      & 18282-0951  & 18:31:01.7 & -09:49:01.4 & 0.10 & 23.90 & 8.0 & 14.5 & 17.1 & 0.04 & 10.0 & 14.5 & 12113 &  1.45\\
G22.36+0.07      & 18290-0924  & 18:31:44.1 & -09:22:12.7 & 0.10 & 23.91 & 8.0 & 15.0 & 17.7 & 0.05 & 10.0 & 15.0 & 11319 & 1.43\\
G23.26-0.24      & 18317-0845  & 18:34:31.8 & -08:42:46.8 & 0.12 & 23.88 & 8.1 & 15.6 & 18.1 & 0.07 & 10.0 & 16.2 & 12914 &  1.38\\
G23.44-0.18      & 18319-0834  & 18:34:39.2 & -08:31:32.5 & 0.15 & 23.87 & 8.2 & 15.8 & 18.2 & 0.08 & 11.0 & 16.4 & 12988 &  1.32\\
G23.71-0.20      & 18324-0820  & 18:35:12.4 & -08:17:39.6 & 0.13 & 23.87 & 8.0 & 15.1 & 17.7 & 0.06 & 10.5 & 15.9 & 11840 &  1.35\\
G24.79+0.08      & 18335-0711a & 18:36:12.3 & -07:12:11.1 & 0.11 & 24.14 & 9.6 & 16.0 & 18.4 & 0.07 & 11.0 & 16.5 & 16342 &  1.35\\
G24.85+0.09      & 18335-0711b & 18:36:18.4 & -07:08:52.1 & 0.12 & 23.85 & 7.9 & 15.3 & 17.7 & 0.06 & 11.0 & 15.9 & 10667 &  1.42\\
G25.65+1.05      & 18316-0602  & 18:34:20.9 & -05:59:40.4 & 0.14 & 23.91 & 8.1 & 16.0 & 18.3 & 0.08 & 11.0 & 16.5 & 11660 &  1.42\\
G25.71+0.04      & 18353-0628  & 18:38:03.1 & -06:24:14.7 & 0.12 & 23.85 & 8.2 & 15.0 & 17.5 & 0.05 & 11.0 & 15.7 & 13210 &  1.35\\
G25.83-0.18      & 18361-0627  & 18:39:03.6 & -06:24:09.9 & 0.16 & 23.88 & 8.0 & 15.1 & 17.7 & 0.06 & 11.5 & 15.9 & 16948 &  1.27\\
G28.15+0.00      & 18403-0417a & 18:42:42.2 & -04:15:31.9 & 0.12 & 23.83 & 6.5 & 14.3 & 17.8 & 0.03 & 10.0 & 14.9 & 6051 &  2.61\\
G28.20-0.05      & 23.597-0417b & 18:42:58.1 & -04:13:56.0 & 0.10 & 23.60 & 8.1 & 15.4 & 17.6 & 0.07 & 10.0 & 15.9 & 12790 &  1.29\\
G28.28-0.36      & 18416-0420b & 18:44:13.3 & -04:18:03.0 & 0.11 & 23.91 & 6.9 & 14.6 & 18.1 & 0.05 & 10.0 & 15.1 & 8818 & 2.35\\
G28.83-0.25      & 18421-0348b & 18:44:51.1 & -03:45:48.2 & 0.11 & 23.92 & 6.5 & 14.7 & 17.9 & 0.04 & 10.0 & 15.0 & 8751 &  2.15\\
G28.85-0.23      & 18421-0348a & 18:44:47.8 & -03:44:16.8 & 0.11 & 23.94 & 7.3 & 14.3 & 17.3 & 0.03 & 10.5 & 14.5 & 6794 &  2.16\\
G29.87-0.04      & 18434-0242b & 18:46:00.0 & -02:44:58.1 & 0.10 & 23.87 & 6.6 & 14.7 & 17.7 & 0.03 & 11.5 & 14.9 & 7586 &  2.00\\
G29.96-0.02      & 18434-0242c & 18:46:04.8 & -02:39:19.7 & 0.10 & 23.93 & 7.3 & 15.2 & 18.0 & 0.04 & 11.5 & 15.2 & 8356 &  2.05\\
G29.98-0.04      & 18434-0242a & 18:46:12.1 & -02:38:58.1 & 0.14 & 23.88 & 7.3 & 15.1 & 18.0 & 0.04 & 12.0 & 15.1 & 8381 &  2.09\\ 
G30.59-0.04      & 18443-0210  & 18:47:18.6 & -02:06:07.0 & 0.16 & 23.85 & 7.7 & 15.3 & 18.1 & 0.05 & 11.0 & 15.2 & 9618 &  1.74\\
G30.71-0.06      & 18450-0200d & 18:47:36.5 & -02:00:31.3 & 0.16 & 23.87 & 7.8 & 15.4 & 18.0 & 0.05 & 11.0 & 15.2 & 9455 & 1.74\\
G30.78+0.23      & 18440-0148c & 18:46:41.5 & -01:48:32.2 & 0.11 & 23.92 & 7.9 & 15.4 & 17.9 & 0.05 & 10.0 & 16.0 & 9226 & 1.56\\
G30.79+0.20      & 18440-0148b & 18:46:48.1 & -01:48:46.0 & 0.11 & 23.93 & 8.1 & 15.9 & 18.4 & 0.07 & 10.0 & 16.5 & 12339 & 1.61\\
G30.82-0.05      & 18450-0200a & 18:47:46.5 & -01:54:16.8 & 0.13 & 23.86 & 8.0 & 15.5 & 18.0 & 0.06 & 10.0 & 16.1 & 11487 & 1.43\\
G30.82+0.28      & 18440-0148a & 18:46:36.1 & -01:45:18.2 & 0.20 & 23.91 & 8.2 & 15.5 & 17.9 & 0.05 & 10.0 & 16.1 & 11223 &  1.43\\
G30.90+0.16      & 18446-0150  & 18:47:09.2 & -01:44:09.4 & 0.11 & 23.81 & 8.1 & 15.5 & 18.1 & 0.05 & 11.0 & 16.1 & 10363 & 1.47\\
G31.06+0.09      & 18452-0141  & 18:47:41.2 & -01:37:21.3 & 0.12 & 23.83 & 8.1 & 15.6 & 18.1 & 0.06 & 11.0 & 16.2 & 14469 &  1.33\\
G31.28+0.06      & 18456-0129  & 18:48:12.4 & -01:26:22.6 & 0.10 & 23.89 & 8.0 & 15.6 & 18.2 & 0.05 & 10.0 & 16.3 & 11903 &  1.55\\
G31.41+0.31      & 18449-0115  & 18:47:34.3 & -01:12:47.1 & 0.10 & 23.93 & 8.2 & 15.4 & 17.9 & 0.05 & 11.0 & 16.2 & 12839 & 1.49\\ \hline

    \end{tabular}
    
    \end{tiny}
  \end{table*}

\begin{table}
\begin{scriptsize}
  \begin{center}
  \caption{Example extract from the first five lines of the IRIS2
    Ks-band (2.14\,$\mu$m) point source catalogue towards a sample
    region. The first two columns give the source right ascension and
    declination in decimal degrees. Columns 3 and 4 give the magnitude
    and magnitude uncertainty. The full tables will be available on
    line.}
  \label{tab:i2ks_psc_examp}

  \begin{tabular}{|c|c|c|c|}\hline\hline 
R.A.      &  Dec.     & Mag.    & $\Delta$Mag. \\ \hline 
272.1961  & -20.1837  & 16.133  & 0.065 \\
272.2225  & -20.1838  & 17.37  & 0.144 \\
272.2219  & -20.1837  & 17.108  & 0.119 \\
272.2214  & -20.1837  & 17.31  & 0.125 \\
272.2209  & -20.1833  & 15.706  & 0.041 \\ \hline

  \end{tabular}
  
  \end{center}
  \vspace{-2mm}
\end{scriptsize}
\end{table}

%------------------------------------------------------------

\newpage
\clearpage

\begin{table*}
\begin{tiny}
  \begin{center}
  \caption{Example extract from the first five lines of the online
    catalogue of the sources matched between the IRIS2 Ks-band
    (2.14\,$\mu$m) images and the GLIMPSE point source catalogue. The
    first column gives the GLIMPSE ID number. The second and third
    columns give the source right ascension and declination in decimal
    degrees. Columns 4 and 5 give the magnitude and magnitude error of
    the sources at Ks band. Columns 6 through 13 give the magnitude
    and magnitude error of the sources at 3.6, 4.5, 5.8 and
    8.0\,$\mu$m from the GLIMPSE point source catalogue. In this
    example table, the magnitudes and errors have been rounded to 2
    decimal places for clarity. A value of `null' means no source was
    detected at that wavelength. The full table is available
    online. Note that these PSCs \emph{ONLY} include sources matched
    in both the GLIMPSE and IRIS2 Ks catalogues. Photometry for
    sources detected only in the IRIS Ks PSC (likely fainter
    foreground stars) or GLIMPSE PSC (either due to being too faint,
    or possibly in saturated/confused regions) can be extracted from
    these individual PSCs. Finally, photometry for sources in the
    IRIS2 Ks PSC brighter than the minimum reliable magnitude reported
    in column 11 of Table~\ref{tab:daophot_pars}, should be extracted
    from the 2MASS PSC.}
  \label{tab:i2ks_glimpse_psc_examp}

%  \vspace{1cm}
  \begin{tabular}{|c|c|c|c|c|c|c|c|c|c|c|c|c|c|}\hline\hline 
   ID & RA J2000  & Dec. J2000 & Ks   & $\Delta$Ks & 3.6 & $\Delta$3.6 & 4.5 & $\Delta$4.5 & 5.8 & $\Delta$5.8 & 8.0 & $\Delta$8.0 \\
   GLIMPSE & (degrees) & (degrees)  & (mag) & (mag)      &(mag)& (mag)       &(mag)&(mag)        &(mag)&(mag)        &(mag)& (mag)      \\ \hline

SSTGLMC\_G010.2102-00.1517 & 272.1905 & -20.17984 & 12.86 & 0.02 & 11.90 & 0.10 & 11.72 & 0.09 & 11.79 & 0.25 & null & null \\
SSTGLMC\_G010.2213-00.1721 & 272.2152 & -20.18006 & 14.02 & 0.02 & 11.86 & 0.07 & 11.43 & 0.07 & 10.10 & 0.13 & 10.83 & 0.01 \\
SSTGLMC\_G010.2238-00.1767 & 272.2208 & -20.1801  & 12.02 & 0.04 & 11.51 & 0.08 & 11.54 & 0.09 & 11.35 & 0.15 & 10.53 & 0.16 \\
SSTGLMC\_G010.2118-00.1511 & 272.1907 & -20.17816 & 13.25 & 0.02 & 12.04 & 0.08 & 12.11 & 0.13 & 11.71 & 0.19 & null & null \\
SSTGLMC\_G010.2431-00.2087 & 272.2606 & -20.17873 & 14.06 & 0.02 & 12.42 & 0.08 & 12.06 & 0.10 & null & null & null & null \\ \hline

  \end{tabular}
  
  \end{center}
  \vspace{-2mm}
\end{tiny}
\end{table*}

%------------------------------------------------------------
\clearpage
\newpage
\subsection{Creating the point source catalogues}
\label{sub:creating_PSCs}

This section describes how the DAOPHOT package (digi.daophot) in IRAF
was used to automatically find and perform photometry on stars within
the IRIS2 fields and hence create the point source
catalogues. Description of co-ordinate frame calibration, photometric
calibration and catalogue matching are found in
Section~\ref{sec:iris2_obs}.

\paragraph{Set initial fitting parameters}
\label{subsub:daophotpars}
As three seperate detector arrays were used over the course of the
observations care was taken to measure the array specific input
parameters (eg readout noise, gain, saturation value etc.) for each
image.  The image characteristics used by the star finding and
photometry tasks were calculated using \emph{daoedit} to pick out 10
bright, non-saturated, isolated stars with as little surrounding
extended emission as possible. By fitting a radial profile to the
emission, this task then calculates the FWHM, SKY (median) and
$\sigma_{SKY}$ for each of the stars. After calculating a sensible
average value for each of these parameters from stars with good
profiles, the values are then entered into several `pars' tasks (which
act as inputs into the finding/photometry tasks) as follows,

\begin{itemize}
\item epar datapars:\\fwhmpsf = FWHM;\\sigma = $\sigma_{SKY}$;\\
datamin
= SKY-(6$\times$$\sigma_{SKY}$);\\datamax  = ``saturation value''
(must be less than max value ie max `good' data value);
\item epar centerpars:\\ cbox = 2$\times$FWHM or 5 (whichever is larger)
\item epar fitskypars:\\ salgori = mode (for crowded images);\\ annulus =
4xFWHM; \\ dannulus = 3.5$\times$FWHM
\item epar photpars:\\ aperture = 1$\times$FWHM
\item epar daopars:\\ psfrad = (4$\times$FWHM)+1;\\ fitrad = 1$\times$FWHM
or 3 (whichever is larger)
\end{itemize}

\paragraph{Intial star list}
\label{subsub:daophot_init_list}
After calculating the image parameters, the initial (first-pass) star
list was created using the \emph{daofind} task to run the search
finding algorithm. This task was run several times using different
detection thresholds between 3 and 5$\sigma$ and compared by eye to
maximise finding fainter stars and excluding noise.

\paragraph{Initial photometry}
\label{subsub:daophot_init_phot}
Photometry of the stars in the first-pass list was calculated using
\emph{daofind}. Stars dumped by the task due to poor fits were
manually checked to ensure no real stars were dropped. In practice,
the dumped stars were either low signal noise spikes or saturated
stars.

\paragraph{Calculating the point spread function}
\label{subsub:daophot_psfcalc}
The task \emph{pstselect} was used to select 10 bright, non-saturated,
isolated stars across the image to calculate the PSF. A characteristic
PSF for the image was then calculated from these PSF stars. The
robustness of the PSF was checked by subtracting the characteristic
PSF from the 10 selected stars and their nearest neighbours and
inspecting the residuals. In an iterative process, bad PSF stars were
removed and the procedure repeated until good residuals were seen
across the image meaning a realiable characteristic PSF for that image
had been obtained.

\paragraph{Extracting the stars and fitting world coords}
\label{subsub:daophot_star_extract}
Once the optimal image parameters and characteristic PSF had been
obtained, the PSF model was fit to all the stars using \emph{allstar}.
The residuals after the PSF subtraction were carefully checked to
ensure both bright and faint stars across the image had been
successfully subtracted. The star coordinates were converted from
pixels to world coordiates using \emph{wcsctran}.

%------------------------------
%new, decent images
%\input figs/IRIS2_Ks_figs

\end{document}